\documentclass[preprint,aps,nofootinbib]{revtex4}
\usepackage{graphicx}

\begin{document}


\title{
Implications of the Higgs Discovery in the MSSM Golden Region
}
\vspace*{2cm}

\author{
\vspace{1cm}
Ian Low and Shashank Shalgar
}
\affiliation{
\vspace*{.5cm}
\mbox{Theory Group, HEP Division, Argonne National Laboratory, Argonne, IL 60439} \\
\mbox{Department of Physics and Astronomy, Northwestern University, Evanston IL 60208}
\vspace*{1cm}
}

\begin{abstract}
\vspace*{0.5cm}
If the lightest CP-even Higgs boson in the MSSM 
is discovered at the LHC, two measurements could be made simultaneously: the Higgs
mass $m_h$ and the event rate $B\sigma(gg\to h\to \gamma\gamma)$. We study to what extent the combination of these two measurements
would allow us to extract parameters in the stop mass matrix, including the off-diagonal mixing term, with a focus on the MSSM golden
region where the stops are light and the mixing is large. Even though both the production cross-section and the decay amplitude
are not sensitive to supersymmetric parameters outside of the stop sector, the branching ratio depends on the total decay
width, which is dominated by the Higgs decay to $b$ quarks and sensitive to both the 
 pseudo-scalar mass $m_A$ and the supersymmetric Higgs mass $\mu$. In the end we find $m_A$ is an important input
in extracting the stop mass parameters, while a fair estimate of the off-diagonal mixing term could be obtained
without prior knowledge of $\mu$.

\end{abstract}

\maketitle

\section{Introduction}
Supersymmetry (SUSY) is widely considered as the leading proposal to address the hierarchy
problem in the standard model. The simplest realization of SUSY, 
the mininal supersymmetric
standard model (MSSM), is without a doubt the most popular extension of the 
standard model in the last three decades. Although a 
theoretically very appealing concept, SUSY has remained elusive in experimental searches
 despite tremendous amount of efforts. The non-discovery of any new particles or the Higgs boson
to date introduces a certain amount of fine-tuning in the electroweak sector of MSSM 
\cite{Kitano:2005wc, Giudice:2006sn} and
creates a somewhat awkward situation given SUSY's promise to solve the electroweak 
hierarchy problem.

There is, however, a particular region in the parameter space of MSSM where the fine-tuning
in the Higgs mass is reduced. It is the region where the overall stop mass scale
is light at several hundreds GeV and the mixing in the stop sector is maximized due to a large
trilinear soft $A_t$ term, which implies two stop mass eigenstates below 1 TeV with a mass 
splitting in the order of a few hundreds GeV. Such a region is dubbed the ``golden region''
of MSSM by Perelstein and Spethmann in Ref.~\cite{Perelstein:2007nx} because it satisfies 
the experimental constraints,
including the bound of 114 GeV on the Higgs mass, and minimizes the unnaturalness
in the MSSM. 
Obviously implications of the MSSM golden region on the possible 
collider signatures at the Large Hadron Collider (LHC) deserve
detailed studies if one wish to take the naturalness argument seriously. Previous 
works on the phenomenology of MSSM golden region can be found in 
Refs.~\cite{Kitano:2006gv,Dermisek:2007fi, Perelstein:2007nx, Essig:2007vq,
Essig:2007kh,Kasahara:2008up,Cho:2008rp}. 


In this work we wish to study the implication of discovering the lightest CP-even Higgs in 
the MSSM golden region. Since the Higgs sector of the MSSM contains two Higgs doublets, after
electroweak symmetry breaking there are five scalar Higgs bosons remaining: a pair of charged
Higgs $H^\pm$, one CP-odd Higgs $A$, two CP-even Higgses $h$ and $H$ where $H$ is defined to
be the heavier one. Moreover,
it is found that, after including radiative corrections, there is an upper bound on 
the mass of the lightest CP-even Higgs at roughly $130-140$ GeV. 
(For a recent review, see Ref.~\cite{Djouadi:2005gj}.) The importance of such a bound lies
in the observation that a Higgs boson in this mass range can be discovered at the LHC only 
through its decay into $\gamma\gamma$, which is a loop-induced process, because its mass is below
the $W^+W^-$ and $ZZ$ thresholds; decays into $b\bar{b}$ pair are swamped by the background. 
On the other hand, the dominant production mechanism
of the neutral Higgs boson at the LHC is the gluon fusion process \cite{Djouadi:2005gi}, also
a loop induced process. Since both production and decay channels are loop induced, 
and hence very sensitive to effects of new physics if any,
the process $gg\to h \to \gamma\gamma$ is the natural playground to look for deviations from standard
model predictions.
 
Within MSSM effects of supersymmetric particles in 
 the production cross-section $\sigma(gg\to h)$ and the partial decay width $\Gamma(h \to \gamma\gamma)$ 
have been studied 
previously \cite{Djouadi:1996pb, Djouadi:1998az}. It is found that for both amplitudes the
leading corrections come from
the stop sector and, to a lesser extent, charged SUSY particles such as the charginos. However, 
it is important to emphasize that, once the Higgs is discovered in the process $gg\to h \to \gamma\gamma$, 
two experimental measurements can be made at the same time: the event 
rate $\sigma(gg\to h)\times Br(h\to \gamma\gamma)$
and the Higgs mass $m_h$. One should therefore make use of the information from 
both measurements in trying to extract 
parameters of the MSSM. Indeed, Ref.~\cite{Dermisek:2007fi} showed that it is possible to
obtain a fairly good estimate on the 
trilinear soft $A_t$ term in the stop sector, which is otherwise difficult to measure, 
if one combines knowledge on 
the production cross section in the gluon fusion channel $\sigma(gg\to h)$ and $m_h$. 
One limitation of the proposal in 
\cite{Dermisek:2007fi} is that the production cross section $\sigma(gg\to h)$ is not directly observable 
in collider experiments\footnote{There are, nevertheless, ways to extract this cross section 
at the LHC with an uncertainty in the order of
$30-40$\% \cite{Duhrssen:2004cv}, even though a recent study suggests a smaller uncertainty 
may be possible \cite{Anastasiou:2005pd}.};
only the product $\sigma(gg\to h)\times Br(h\to \gamma\gamma)$ is. Therefore, in this 
note we will study implications of combining
two measurements, $\sigma \times Br$ and $m_h$,
 in $gg\to h \to \gamma\gamma$ with an emphasis on the golden region of the MSSM parameter space.

\section{Higgs couplings to $gg$ and $\gamma\gamma$ in the MSSM}

In this section we discuss the effects of superparticles on the Higgs coupling to two gluons
and two photons. In particular, we will work in the so-called decoupling limit \cite{Haber:1995be} where 
the pseudo-scalar Higgs is much heavier than the $Z$ boson: $m_A \gg m_Z$. There are two reasons for considering
the decoupling limit. 
The first is the lightest CP-even Higgs quickly reaches its maximal possible mass, which
is phenomenologically desirable given the null result for Higgs searches at both LEP and Tevatron. The second
reason is the tree-level couplings of $h$ in the MSSM 
to standard model fermions and gauge bosons approach the values for a standard model Higgs, and the two 
become almost indistinguishable. Therefore any deviations in the event rate from the standard model would
come from effects of superparticles running in the loop. It is also worth mentioning that in this case
all other Higgs bosons in the MSSM are roughly degenerate $m_A\approx m_{H} \approx m_{H^\pm}$.
In practice, the decoupling limit is quickly reached for moderately large $\tan \beta \agt 10$ whenever
$m_A$ gets larger than the maximal possible value for $h$: $m_A \agt m_h^{\rm max}$. For small $\tan\beta$
the decoupling limit is reached when $m_A \agt 300$ GeV. However, we will see later that Higgs couplings
to vector bosons approach the decoupling limit much faster than Higgs couplings to fermions, which has important
implications in the present study.

Analytic expressions for a CP-even scalar Higgs decaying into two gluons and two photons within 
the standard model were obtained long ago in Refs.~\cite{Ellis:1975ap,Georgi:1977gs,Shifman:1979eb}. The 
loop induced amplitude $h\to gg$ is mediated by the top quark, which has the largest coupling to the Higgs among
the fermions, whereas $h\to \gamma\gamma$ has the leading contribution from the $W$ boson loop with the top 
quark as the subleading contribution. In the MSSM expressions for effects of
 superpartners can be found in Ref.~\cite{Gunion:1989we}. In $h\to gg$ the top squark gives the dominant 
supersymmetric contribution, while all the charged superparticles, such as the sfermions, charged Higgs 
scalars, and the charginos, now contribute to $h\to \gamma\gamma$ as well. More explicitly,
\begin{eqnarray}
\Gamma(h \to gg) &=& \frac{G_F \alpha_s^2 m_h^3}{36 \sqrt{2}\pi^3} 
       \left| N_{c}\, Q_{t}^{2}\,  g_{htt}\, A_{\frac12}^h(\tau_t) +  {\cal A}^{gg}
     \right|^2 , \\
\Gamma(h\rightarrow\gamma\gamma)&=&\frac{G_{F}\alpha^{2}m_{h}^{3}}{128\sqrt{2}\pi}
\left| N_{c}\, Q_{t}^{2}\, g_{htt}\, A_{1/2}^{h}(\tau_{t})+
g_{hWW}\, A_{1}^{h}(\tau_{W}) + {\cal A}^{\gamma\gamma}\right|^2,
\end{eqnarray}
where $g_{htt}$ and $g_{hWW}$ are the coupling of $h$ to the top quark and the $W$ boson, respectively.
Moreover $\tau_i = m_h^2/(4m_i^2)$ and the form factors are
\begin{eqnarray}
A_{\frac12}^h(\tau) &=& \frac2{\tau^2} [ \tau + (\tau - 1)f(\tau)]\ , \\
A_0^h(\tau) &=& -\frac{1}{\tau^2}\left[ \tau - f(\tau) \right] \ , \\
A_1^h(\tau) &=&  -\frac{1}{\tau^2}[2\tau^2+3\tau+3(2\tau-1)f(\tau) ] \ , \\
\label{ftau}
f(\tau) &=& \left\{ \begin{array}{lc}
       \displaystyle     \arcsin^2 \sqrt{\tau} & \quad \tau \le 1 \\
  \displaystyle -\frac14 \left[ \log 
    \frac{1+\sqrt{1-\tau^{-1}}}{1-\sqrt{1-\tau^{-1}}} - i \pi \right]^2 
      & \quad \tau > 1 
         \end{array} \right. .
\end{eqnarray}
The supersymmetric contributions ${\cal A}^{gg}$ and ${\cal A}^{\gamma\gamma}$ are
\begin{eqnarray}
{\cal A}^{gg} &=&  \sum_{i} N_{c}\, Q_{t}^{2}\, g_{h\tilde{t}_i\tilde{t}_i}\, \frac{m_Z^2}{m_{\tilde{t}_i}^2}\,
                 A_{0}^h(\tau_{\tilde{t}_i})\ , \\
{\cal A}^{\gamma\gamma} &=&  
 g_{hH^{+}H^{-}}\, \frac{m_{W}^{2}}{m^{2}_{H^{\pm}}} \, A_{0}^{h}(\tau_{H^{\pm}}) +
\sum_f N_c Q_f^2\, g_{h\tilde{f}\tilde{f}}\, \frac{m_Z^2}{m^2_{\tilde{f}}}\, A_0^h(\tau_{\tilde{f}})
     + \nonumber \\
&& \quad \quad \quad \sum_i g_{h\chi_i^+\chi_i^-}\, \frac{m_W}{m_{\chi_i}}\, A_{\frac12}^h(\tau_{\chi_i}) 
\end{eqnarray}
Explicit forms for the couplings of $h$ to superparticles can be found in 
Refs.~\cite{Djouadi:2005gj, Gunion:1989we}. We will, however, write down 
$g_{h\tilde{t}_i\tilde{t}_i}$ since the stop contribution is the focus of present study.
When normalizing to $2m_Z^2 (\sqrt{2}G_F)^{1/2}$,
\begin{eqnarray}
g_{h\tilde{t}_1\tilde{t}_1} &=&   \cos 2\beta
   \left(\frac12 \cos^2 \theta_t-\frac23 s_w^2 \cos 2\theta_t\right)  
           +  \frac{m_t^2}{m_Z^2} - \frac12 \frac{m_t X_t}{m_Z^2}  \sin 2\theta_t , \\
g_{h\tilde{t}_2\tilde{t}_2} &=&   \cos 2\beta
   \left(\frac12 \sin^2 \theta_t+\frac23 s_w^2 \cos 2\theta_t\right) 
           +  \frac{m_t^2}{m_Z^2} + \frac12 \frac{m_t X_t}{m_Z^2}  \sin 2\theta_t .
\end{eqnarray}
In the above $s_w$ is the sine of Weinberg angle and $\theta_t$ is the mixing angle in the stop sector. 
The mixing parameter $X_t$ is the off-diagonal entry in the stop mass matrix
\begin{equation}
\label{stopmass}
M^2_{\tilde{t}} = \left( \begin{array}{cc} 
   m^2_{\tilde{t}_L} + m_t^2 + D_L^t & m_t X_t \\
    m_t X_t &  m^2_{\tilde{t}_R} + m_t^2 + D_R^t
                         \end{array} \right),
\end{equation}
where
\begin{eqnarray}
D_L^t &=& \left(\frac12-\frac23 s_w^2\right) m_Z^2 \cos 2\beta , \\
D_R^t &=&  \frac23 s_w^2 m_Z^2 \cos 2\beta , \\
X_t &=& A_t - \frac{\mu}{\tan\beta} .
\end{eqnarray}

In the limit of heavy loops masses $\tau_i \ll 1$, the 
form factors approach the asymptotic values
\begin{equation}
\label{asymp}
A_0^h \to - \frac13\ , \quad A_{\frac12}^h \to -\frac43\ ,\quad A_{1}^h \to +7 \ .
\end{equation}
Note that standard model contributions due to the top quark and $W$ boson loops are finite in 
the asymptotic limit, whereas the supersymmetric contributions are suppressed by large loop masses.
One exception is the stop contribution in the MSSM golden region when the stops are light and the mixing
parameter $X_t$ is large. In this case the stop mixing angle is maximal, $|\sin 2\theta_t|\approx 1$, and
the $g_{h\tilde{t}_i\tilde{t}_i}$ coupling is strongly enhanced. Thus the stop correction in the Higgs production
and decay amplitudes is the dominant one and very pronounced, despite being a loop effect.
The particle giving the second largest effect is the chargino, not only because its effect decouples like 
$1/m_{\tilde{\chi}}$ when others decouple like $1/m^2$, as can be seen in ${\cal A}^{\gamma\gamma}$,
but also because of the larger asympotic value of $A_{1/2}^h$ than $A_0^h$.

For the gluon fusion production, it has been shown that the stop could have an order unity effect
when comparing with the standard model rate, especially in the region of light stops and large mixing
$X_t$ \cite{Dermisek:2007fi,Djouadi:1998az}, which is exactly the MSSM golden region. As for the 
Higgs decay into two photons, Ref.~\cite{Djouadi:1996pb} showed that the stop could have a sizable
effect, in the order of 10\% level, in the MSSM golden region. The effect is less dramatic than in the gluon
fusion rate because the $W$ boson loop is the dominant contribution in the decay into two photons, as can be
seen from Eq.~(\ref{asymp}), which shows $A_1^h$ has the largest asymptotic value. 
Moreover, among all the charged superparticles only the chargino loop could have a significant effect again in 
the order of 10\% when the chargino is as light as 100 GeV. For chargino masses above 250 GeV, the deviation
is less than 8\% through out the entire MSSM parameter space \cite{Djouadi:1996pb}. The
direct search limit of charginos from 
LEP is at 103 GeV, which is quite robust and independent of model assumptions, whereas searches at Tevatron
result in a lower bound of 145 GeV in a specific model choice \cite{Yao:2006px}. In Section IV we will 
demonstrate that throughout the MSSM golden region the chargino has a small effect, when considering the ratio of the 
event rate $\sigma(gg\to h)\times Br(h\to \gamma\gamma)$ in the MSSM over the SM, once its mass is heavier than 200 GeV.

In addition to focusing on the region of MSSM parameter space where the stops are light and the mixing is large,
we would follow Ref.~\cite{Dermisek:2007fi} and concentrate on the following choice of parameters:
\begin{itemize}

\item $10 \; \lesssim \; \tan \beta \; \lesssim m_t/m_b$,

\item $|m_b \, \mu \tan \beta| \; \lesssim \; m^2_{\tilde{b}_L}, \ m^2_{\tilde{b}_R}$,

\end{itemize}
The main reason for doing so is to avoid the region where the sbottom contribution could be 
significant in the loop induced processes \cite{Djouadi:1998az} as well as the Higgs
mass \cite{Brignole:2002bz}. Furthermore, the mass of the lightest CP-even Higgs is insensitive 
to $\tan\beta$ once $\tan\beta \agt 10$. Since the bottom quark mass is very small $m_b \sim 5$ GeV,
the above choice covers a very substantial region of MSSM parameter space. 

Another important comment is related to the observation made in Ref.~\cite{Dermisek:2007fi} where it was
shown that
 the gluon fusion production rate depends on only two out of the three parameters in  the stop mass matrix 
in Eq.~(\ref{stopmass}).
If we define the following parameters
\begin{equation}
m_{\tilde{t}}^2 = \frac{m_{\tilde{t}_L}^2+m_{\tilde{t}_R}^2}2, \quad 
r=\frac{m_{\tilde{t}_L}^2-m_{\tilde{t}_R}^2}{m_{\tilde{t}_L}^2+m_{\tilde{t}_R}^2},
\end{equation}
then the cross section $\sigma(gg\to h)$ only depends on $m_{\tilde{t}}^2$ and $X_t$ mostly; the dependence on $r$ is minimal when
$|r|\alt 0.4$. 
\begin{figure}[t]
\includegraphics[scale=0.95,angle=0]{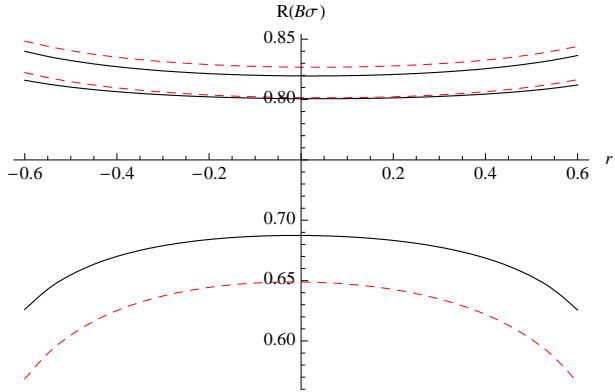}
\caption{\label{fig1}{\it Plot of $R(B\sigma)= B\sigma({\rm MSSM})/B\sigma({\rm SM})$ as a 
function of $r$ for $m_{\tilde{t}}=500$ GeV, $m_A=400$ GeV, $\mu=200$ GeV,and $M_{\rm SUSY}=1$ TeV. The (dark) solid and
(red) dashed lines
are for $\tan\beta=10$ and $40$, respectively. The three sets of curves from top to bottom are 
for $X_t/m_{\tilde{t}}=0, -1,$ and $-2$ 
respectively.}}
\end{figure}
In Fig.~\ref{fig1} we show that this
is the case also in the ratio of the event rate $B\sigma(gg\to h\to \gamma\gamma)$ as well. 
From the figure we see that the variation in the ratio of $B\sigma(gg\to h\to \gamma\gamma)$ is 
less than 10\% for large $X_t$ if $r\alt 0.5$.
For $m_{\tilde{t}}=500$ GeV this translates into $m_{\tilde{t}_L}\sim 600$ GeV and $m_{\tilde{t}_R}\sim 300$ GeV.

It turns out that, as pointed out in Ref~.\cite{Dermisek:2007fi}, the Higgs mass $m_h$ is also sensitive to the same
two parameters in the stop mass matrix. Thus in the end two measurements in $m_h$ and $B\sigma$ could potentially yield
useful information on two mass parameters,
including $X_t$, in the stop sector of the MSSM.

\section{The Higgs Mass in the MSSM}

In this section we very briefly discuss the mass of the lightest CP-even Higgs in the MSSM, as
much of it has been studied extensively in the literature.
The Higgs sector of the MSSM contains two Higgs doublets, $H_u$ and $H_d$, which
couples to the up-type and down-type
quarks separately due to constraints from the anomaly cancellation and the holomorphicity of the 
superpotential. After electroweak symmetry breaking five remaining
 physical states are two CP-even neutral Higgs bosons, $h$ (the lighter one) and $H$
(the heavier one), one CP-odd neutral Higgs boson $A$, and the charged Higgses $H^\pm$. 
In the MSSM there are two free parameters in the Higgs sector, taken to be
$\tan\beta$ and $m_A$, and at tree level 
 one can derive an upper bound on $m_h$ \cite{Djouadi:2005gj}:
\begin{equation}
m_h \le m_Z |\cos 2\beta| \le m_Z = 91.2\ \  {\rm GeV} ,
\end{equation}
which is well below the direct search limit of 114 GeV at LEP.

Therefore, large radiative corrections from superparticles with significant couplings  
to $h$ are required to raise $m_h$ above the LEP bound. Among the new particles in MSSM
the stops have the largest coupling to the lightest CP-even Higgs due to the top Yukawa couplings.
(We are avoiding the region of very large $\tan \beta$ and large $\mu$ where the sbottom couplings could
also be significant, as mentioned in the previous section.)
If we assume for simplicity
$m_{\tilde{t}_R} \simeq m_{\tilde{t}_L} = m_{\tilde{t}}$, the
one-loop correction to $m_h$ is approximately given as \cite{Haber:1996fp,Djouadi:2005gj}
\begin{equation}
\Delta m_h^2 \simeq  \frac{3G_F}{\sqrt{2}\pi^2} m_t^4 \left\{ \log\frac{m_{\tilde{t}}^2}{m_t^2} + \frac{X_t^2}{m_{\tilde{t}}^2}
\left(1-\frac{X_t^2}{12 m_{\tilde{t}}^2} \right) \right\}  ,
\label{eq:mh}
\end{equation}
which grows only logarithmically with the stop mass $m_{\tilde{t}}$. Therefore to lift $m_h$ from $m_Z$ to be
above 114 GeV requires very heavy stops if there is no large mixing. Typically for 
$m_{\tilde{t}_R} \simeq m_{\tilde{t}_L}$ the stop masses need to be very large, 
${\cal O}$(1 TeV), to evade the LEP bound. On the other hand,
the up-type Higgs mass-squared increases quadratically with $m_{\tilde{t}}$,
\begin{equation}
\Delta m_{H_u}^2 \simeq - \frac{3}{8\pi^2} m_{\tilde{t}}^2 \log \frac{\Lambda^2}{m_{\tilde{t}}^2}.
\end{equation}
Thus heavy stop masses at around 1 TeV would lead to large (${\cal O} (m_Z^2/m_{\tilde{t}}^2) \lesssim 1\%$) fine-tuning in 
electroweak symmetry breaking. 
However, the fine-tuning can be reduced if the stop masses could be significantly below 1 TeV while at the same time
keeping $m_h \agt 114$ GeV. This is
possible only  if there is large mixing in the stop sector,
in which case the fine tuning can be reduced to the level of 5\%.
The Higgs mass is maximized for $|X_t/m_{\tilde{t}}|\sim 2$ and 
with this mixing light stops, $m_{\tilde{t}_R} \simeq m_{\tilde{t}_L} \simeq 300$ GeV, 
are sufficient to raise the Higgs mass above the LEP limit. 


\section{Results}

\begin{figure}[t]
\includegraphics[scale=0.8,angle=0]{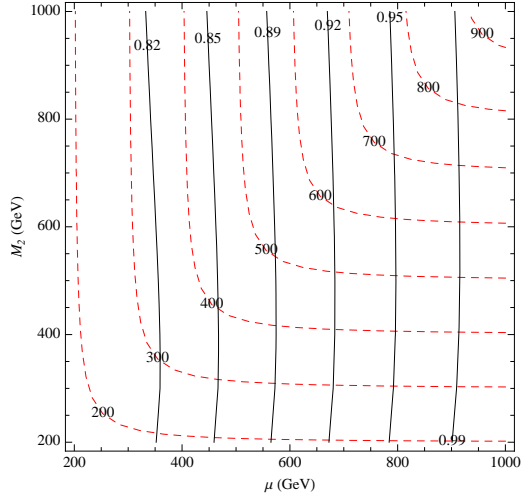}
\caption{\label{fig2}{\it Contour plot of $m_{\chi_1}$ and $R(B\sigma)$ in the $\mu-M_2$ plane.
In this plot $\tan\beta=30$, $M_{SUSY}=1000$ GeV, $m_{\tilde{t}_L}=m_{\tilde{t}_R}= 500$
GeV, and $X_t=-700$ GeV. The (dark) solid lines correspond to the $R(B\sigma)$ contours, whereas the (red) dashed lines are for the $m_{\chi_1}$ contours. 
This set of parameters would result in a Higgs mass slightly above the LEP bound of 114 GeV.}}
\end{figure}

We use the program {\tt FeynHiggs-2.6.4} \cite{Heinemeyer:1998yj, Frank:2006yh} in our numerical analysis with the following relevant 
parameters: $m_t=172.6$ GeV, $m_b=4.2$ GeV, and the pseudo-scalar Higgs mass $m_A=400$ GeV (unless otherwise noted.) 
It is perhaps worth emphasizing that 
{\tt FeynHiggs} employs various approximation schemes in computing the MSSM Higgs production and decay rates, which are extrapolated
from the SM predictions. In this work we use the program mainly to study the dependence on other SUSY parameters and
demonstrate the feasibility of our proposal. Higher-order QCD corrections could be important and should be included if a full-fledged analysis
were performed, which is beyond the scope of the current work. In the following the main observable we consider is the ratio of the event rates in the MSSM over
the standard model: $R(B\sigma)= B\sigma({\rm MSSM})/B\sigma({\rm SM})$. (When we speak of the event rate, we would always have in mind the ratio of 
the event rates!)

We first consider the effect of the chargino mass on $R(B\sigma)$. In MSSM the chargino mass
is determined by the gaugino mass parameter $M_2$ and the supersymmetric Higgs mass parameter $\mu$. 
In Fig.~\ref{fig2} we present a contour plot of $R(B\sigma)$ and the lightest chargino mass $m_{\chi_1}$ in the $\mu-M_2$ plane, in which $m_{\chi_1}$ 
varies from 200 to 900 GeV. We see that contours of constant $R(B\sigma)$ runs somewhat parallel to the $M_2$ axis, 
which implies $R(B\sigma)$ is sensitive to $\mu$ only. Therefore $R(B\sigma)$ depends on the chargino mass
only through its dependence on the $\mu$ parameter. 
For example, let's choose $\mu=900$ GeV and change
 $M_2$ from 200 to 1000 GeV, which results a modification in 
$m_{\chi_1}$ from 200 to 900 GeV. 
The corresponding $R(B\sigma)$ remains roughly constant at 0.99, as can be seen from Fig.~\ref{fig2}, in spite
of a large variation in $m_{\chi_1}$. On the other hand, one could vary $\mu$ while keeping $m_{\chi_1}$ constant,
and $R(B\sigma)$ would change significantly according to $\mu$.
It has been previously observed that
the chargino has a  small effect in the partial width $\Gamma(h\to \gamma\gamma)$ \cite{Djouadi:1996pb}. 
Here we see what is important is the value of the $\mu$ parameter, instead of the chargino mass itself.
If $\mu$ is not too large, then the effect of the chargino is small not
only in the partial width of $h\to \gamma\gamma$, but also in the ratio of
the event rate $B\sigma(gg\to h\to \gamma\gamma)=\sigma(gg\to h)\times Br(h\to \gamma\gamma)$.

On the other hand, we observe in Fig.~\ref{fig2} that $R(B\sigma)$ does have a strong dependence in the supersymmetric Higgs mass parameter $\mu$.
From the discussion above we know this dependence cannot come from the charginos. It turns out the $\mu$ dependence does not arise from effects of 
supersymmetric particles in either the production cross-section $\sigma(gg\to h)$ or the partial decay width $\Gamma(h\to \gamma\gamma)$. Somewhat surprisingly,
the strong $\mu$ dependence resides in the radiative corrections of supersymmetric particles to the bottom Yukawa coupling. The bottom Yukawa coupling controls
the decay amplitude  of $h\to b\bar{b}$, which partial width
 dominates the total decay width of the lightest CP-even Higgs in MSSM. Since the branching ratio is given as
\begin{equation}
Br(h\to\gamma\gamma)=\frac{\Gamma(h\to \gamma\gamma)}{\Gamma_{\rm tot}}\approx \frac{\Gamma(h\to \gamma\gamma)}{\Gamma(h\to b\bar{b})},
\end{equation}
the event rate $B\sigma$ is very sensitive to supersymmetric corrections to $h\to b\bar{b}$.

\begin{figure}[t]
\includegraphics[scale=0.87,angle=270]{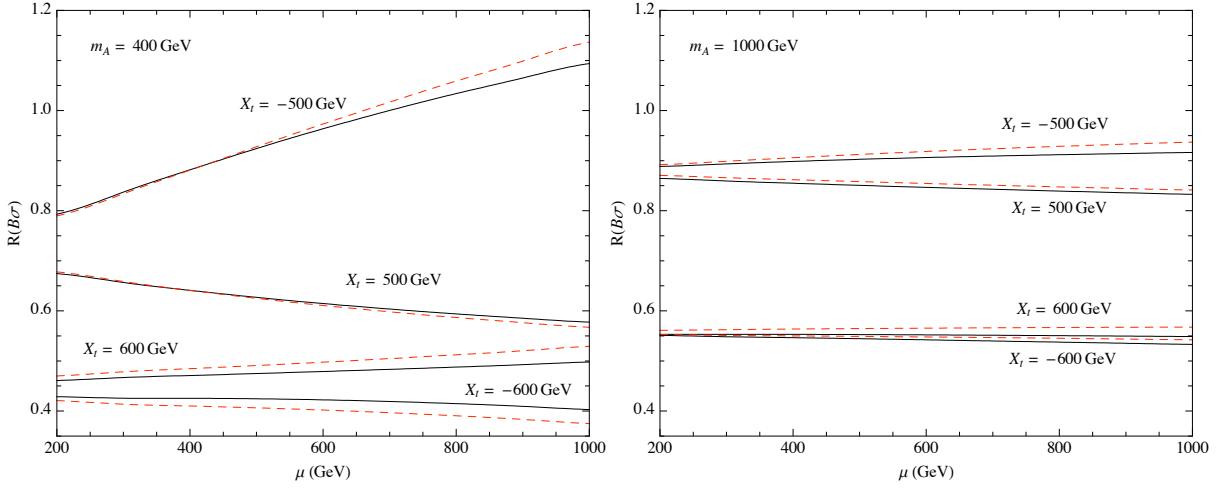}
\caption{\label{fig3}{\it Dependence on $\mu$ in the event rate $R(B\sigma)$. In
these plots $\tan\beta=30$, $M_{SUSY}=500$ (solid lines)  and $1000$ (dashed lines) GeV, 
$m_{\tilde{t}_L}=m_{\tilde{t}_R}= 300$
GeV, and $X_t=\pm 500$ and $\pm 600$ GeV. The two choices of $X_t$ correspond a Higgs mass of 116 and 119 GeV, respectively. The left plot
is for $m_A=$ 400 GeV and the right for $m_A=$ 1000 GeV. In the plots
we have also set the gaugino mass parameter $M_2=M_{SUSY}$.}}
\end{figure}

More specifically, at one-loop the running $b$-quark mass receives contributions from sbottom-gluino loops and stop-charged Higgsino loops\footnote{There 
is also a sbottom-wino loop whose contribution is suppressed by the smallness of the electroweak gauge coupling.}
\cite{Hall:1993gn,Hempfling:1993kv,Carena:1994bv}, which depends on the gluino mass, the stop and sbottom eigenmasses, the supersymmetric Higgs mass
$\mu$ as well as the stop mixing parameter $X_t$. Such corrections become more important when $\tan\beta$ becomes larger, 
which is the regime we are interested in. However, since we are avoiding regions where the sbottoms are light, we find
the sbottom-gaugino loop is numerically unimportant.
In Fig.~\ref{fig3} we single out the $\mu$ dependence of $R(B\sigma)$ for $\tan\beta=30$ and  
some choices of stop masses and $X_t$. The reason that the 
variation is asymmetric in $X_t \to -X_t$ is because the contribution from stop-chargino loop 
is linear in $X_t$ at leading order. So for one sign it adds to the tree-level
result while for the opposite sign it subtracts. Nevertheless, the important observation is that as the mixing parameter $X_t$ becomes larger, the 
variation becomes smaller. In this case, one of the stop mass eigenstates is heavy and suppresses the loop contributions. Therefore if we are only interested
in the MSSM golden region, where $X_t$ is large and $|X_t/m_{\tilde{t}}| \sim 2$, the $\mu$ dependence is reduced.

\begin{figure}[t]
\includegraphics[scale=0.85,angle=0]{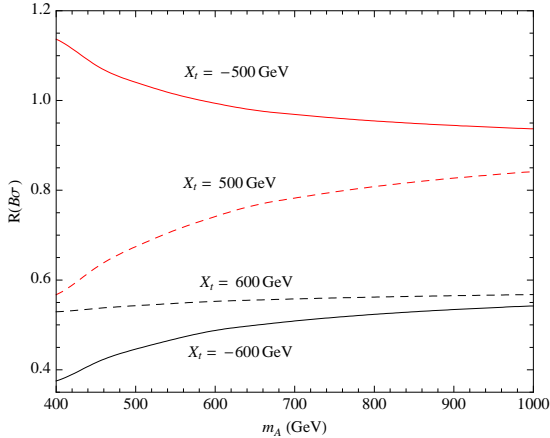}
\caption{\label{fig4}{\it Dependence of $R(B\sigma)$ on $m_A$. In this plot $\tan\beta=30$, $M_{SUSY}=1000$ GeV, 
$m_{\tilde{t}_L}=m_{\tilde{t}_R}= 300$
GeV, and $X_t=\pm 500$ and $\pm 600$ GeV. These two choices of $X_t$ correspond a Higgs mass of 116 and 119 GeV, respectively. In the plot
we have set $\mu=M_2=M_{SUSY}$.}}
\end{figure}

The fact that the total decay width $\Gamma_{\rm tot}\approx \Gamma(h\to b\bar{b})$ enters into $Br(gg\to h\to \gamma\gamma)$ introduces additional 
sensitivity on the mass of the pseudo-scalar Higgs $m_A$, which is absent in either the cross-section $\sigma(gg\to h)$ or the partial decay width 
$\Gamma(h\to \gamma\gamma)$, because the Higgs coupling to bottom quarks depends on $m_A$ in a complicated fashion. When normalizing to 
$im_b/v$, the tree-level Higgs coupling to bottom quarks in MSSM can be written as
\begin{equation}
g_{hbb} = -\frac{\sin{\alpha}}{\cos\beta}; \quad \alpha = \frac12 \arctan \left(\tan2\beta \frac{m_A^2+m_Z^2}{m_A^2-m_Z^2}\right),  
\end{equation}
where $\alpha$ is the mixing angle in the CP-even Higgs sector and $-\pi/2 \le \alpha \le 0$. The usual decoupling regime of MSSM, where the lightest
CP-even Higgs has standard model-like couplings, is obtained in the limit of large $\tan\beta$ and $m_A \gg m_Z$. In this limit the Higgs coupling to
the standard model gauge bosons, $g_{hVV}\propto \sin(\beta-\alpha) \to 1$, approaches that of a standard model Higgs boson. This limit happens quite fast
for moderately heavy $m_A$ and one some times define the decoupling regime as the region where $\sin^2(\beta-\alpha) \ge 0.95$ \cite{Djouadi:2005gj}. However,
because the Higgs coupling to bottom quarks is proportional to  $-{\sin{\alpha}/}{\cos\beta}$, instead of $\sin(\beta-\alpha)$, $g_{hbb}$ approaches its 
standard model value much slower than the $g_{hVV}$ couplings. For example, at $\tan\beta =30$ and $m_A=400$ GeV, we have $\sin^2(\beta-\alpha)=0.99999$ 
whereas $g_{hbb}^2$ is only 0.81, still quite far from the standard model value at unity. In Fig.~\ref{fig4} we plot the $m_A$ dependence of the event
rate $R(B\sigma)$ and one sees that the sensitivity is quite strong in varying $m_A$ from 400 to 1000 GeV, especially for a light Higgs mass $m_h\approx 116$ GeV.

The slow approach to the decoupling limit of the tree-level coupling in $g_{hbb}$ is indicative of the strong $m_A$ dependence in $\Gamma(h\to b\bar{b})$, and hence $\Gamma_{\rm tot}$. 
However, the sensitivity in Fig.~\ref{fig4} cannot be explained by the change in the tree-level coupling alone, and higher-order corrections
play an important role here. Since $m_A$ is taken as one of the input parameters in the Higgs sector of MSSM, it is clear that $m_A$ enters into the
higher-order corrections in a complicated way and its effect cannot be disentangled easily. For example, in addition to the 
supersymmetric loop corrections to the bottom Yukawa couplings mentioned previously, there are also higher-order corrections to the off-diagonal matrix
element of the CP-even Higgs mass matrix \cite{Carena:1998gk, Heinemeyer:2000fa}. We would perhaps only comment that
 all these higher-order corrections have been incorporated in the numerical code {\tt FeynHiggs} \cite{Hahn:2007fq}, which is employed in this study.
It is also worth observing that the dependence of $R(B\sigma)$ on $\mu$ becomes weaker for large pseudo-scalar mass $m_A$, as can be seen in Fig.~\ref{fig3}.

\begin{figure}[t]
\includegraphics[scale=0.08,angle=0]{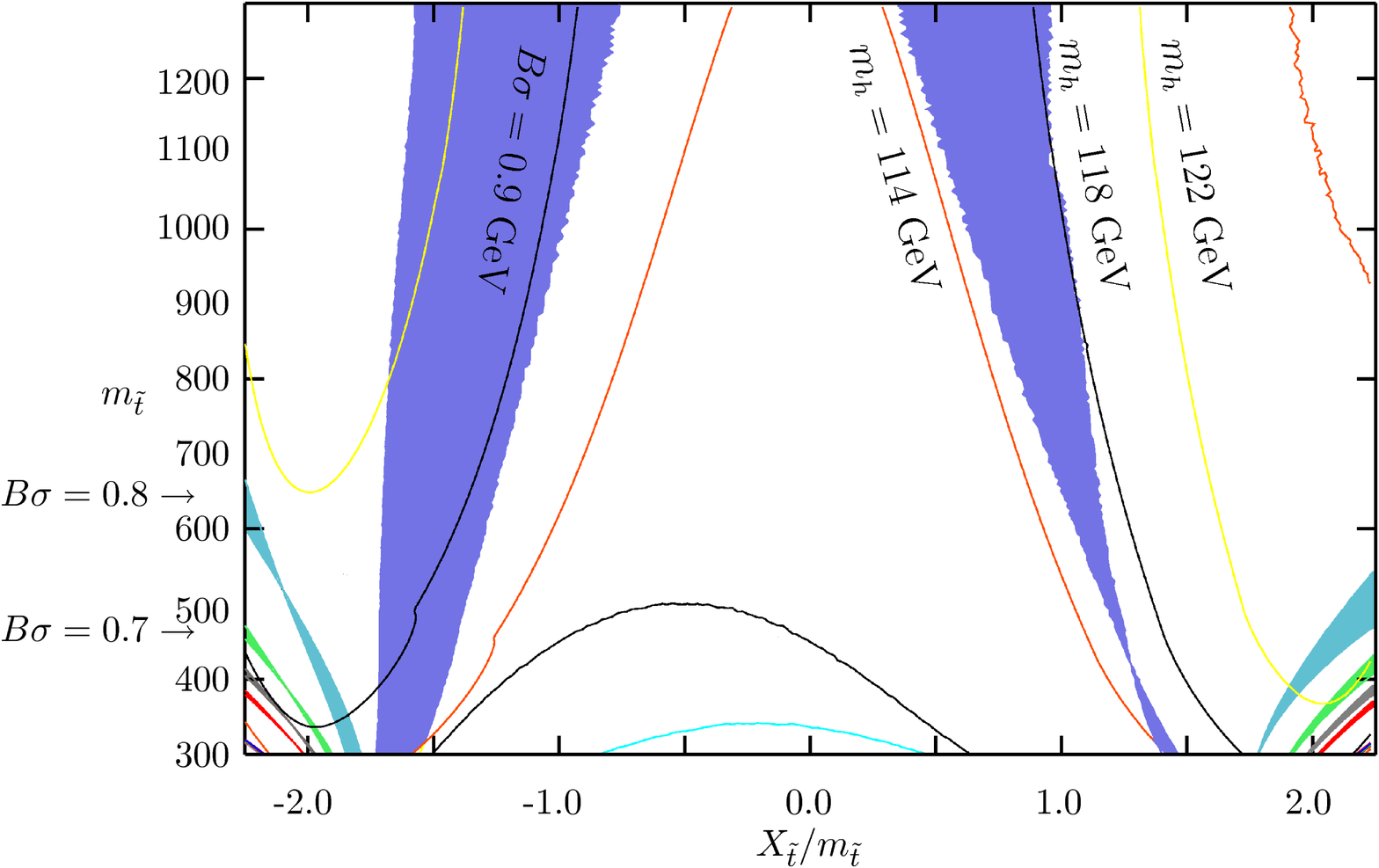}
\caption{\label{fig5}{\it Contours of $R(B\sigma)$ and $m_h$ in the $m_{\tilde{t}}-X_t/m_{\tilde{t}}$ plane. 
This plot is for $\tan\beta=30$, $M_{SUSY}=1000$ GeV, and $m_A$=1000 GeV. The shaded regions correspond to the variation in $R(B\sigma)$ when changing
$\mu$ from 200 to 1000 GeV.}}
\end{figure}

Given the strong $m_A$ dependence and, to a less extent, the $\mu$ dependence in $R(B\sigma)$, it is desirable to have knowledge of these two input parameters
 before one can use both $m_h$ and $B\sigma$
 to extract $m_{\tilde{t}}$ and $X_t$ in the stop sector. This is contrary to the case of utilizing the 
the Higgs production rate in the gluon fusion channel in conjuction with $m_h$, as suggested in \cite{Dermisek:2007fi}, where the sensitivity to SUSY parameters other
than $m_{\tilde{t}}$ and $X_t$ are very weak. In Fig.~\ref{fig5} we present contours of $R(B\sigma)$ and $m_h$ in the $m_{\tilde{t}}-X_t/m_{\tilde{t}}$ plane
for $m_A=1000$ GeV. In the plot we have included the $\mu$ dependence by varying $\mu$ from 200 to 1000 GeV and using the shaded region to
represent the corresponding change in $R(B\sigma)$. We see that in the region where $X_t$ is small and the MSSM fine-tuned, 
the $R(B\sigma)$ contour
runs somewhat parallel to the $m_h$ contours and no useful information on $m_{\tilde{t}}$ and $X_t$ could be extracted. On the other hand, in the
less fine-tuned MSSM golden region where the stops are light and $X_t$ is large, the two contours run almost 
perpendicular to each other and one could
potentially get a fair estimate on the magnitudes of $m_{\tilde{t}}$ and $X_t$. Moreover, in this region of particular interests, $R(B\sigma)$ deviates 
substantially from unity in that the event rate in the MSSM is much smaller than in the standard model, which implies it may take longer time
and more statistics to observe the Higgs at the LHC if indeed the MSSM is realized in nature in a less fine-tuned region of parameter space. In addition, 
the variation due to $\mu$ in $R(B\sigma)$ is also smaller in this particular region.

\begin{figure}[t]
\includegraphics[scale=0.06,angle=0]{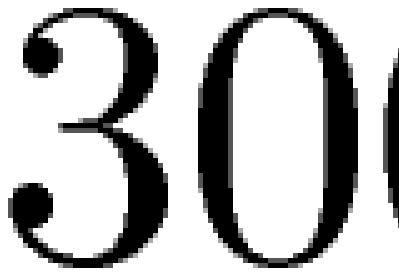}
\includegraphics[scale=0.063,angle=0]{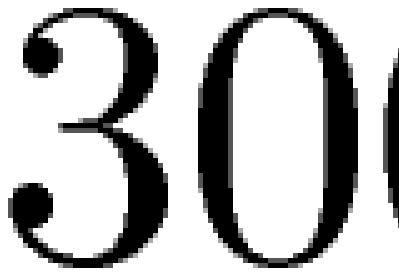}
\caption{\label{fig6}{\it Contours of $R(B\sigma)$ and $m_h$ in the MSSM golden region. In
these plots $\tan\beta=30$ and $M_{SUSY}=1000$ GeV. In the top plot $m_A$=400 GeV and in the bottom plot $m_A=1000$ GeV. 
The shaded regions have the same meaning as in Fig.~\ref{fig5}.}}
\end{figure}

In Fig.~\ref{fig6} we focus in the region where 300 GeV $\le m_{\tilde{t}} \le$ 600 GeV and -2.25 $\le X_t/m_{\tilde{t}} \le $ -1.0 for two different values
of $m_A$ at 400 and 1000 GeV. We have chosen the minus sign for $X_t$ since 
constraints from rare $B$ decays seem to favor negative $X_t$ \cite{Carena:2008ue}.
As commented previously, the variation due to $\mu$ is larger for smaller value of $m_A$. We see that it is important to know
the mass of the pseudo-scalar Higgs $m_A$. But once $m_A$ is known, one could get a fair estimate of $m_{\tilde{t}}$ and $X_t$ even without prior
knowledge of the supersymmetric
Higgs mass $\mu$, especially in the region where $R(B\sigma)\alt 0.8$.

\section{Conclusion}

In this work we studied implications of discovering the lightest CP-even Higgs boson in the MSSM golden region, when measurements on the
Higgs mass $m_h$ and the event rate $B\sigma(gg\to h \to \gamma\gamma)$ are made at the same time. Previously it was suggested that $m_h$ and the Higgs
production cross-section in the gluon fusion channel could be used to extract two parameters, $m_{\tilde{t}}$ and $X_t$, in the stop sectors, which are
important to understand the degree of fine-tuning in the MSSM. We find that in the case of the event rate there are additional sensitivities on the 
pseudo-scalar mass $m_A$ and the supersymmetric Higgs mass $\mu$. It turns out that both sensitivities result from the Higgs coupling to bottom quarks, which
partial decay width dominates the total decay width of the lightest CP-even Higgs boson in the MSSM. The fact that the branching ratio $Br(h\to \gamma\gamma)$ 
depends on the total width introduces sizable dependence of $B\sigma$ on $m_A$ and $\mu$, even though neither the production cross-section $\sigma(gg\to h)$ nor
the partial decay width $\Gamma(h\to \gamma\gamma)$ is sensitive to these two parameters. We also find that, in the MSSM golden region where the stops are
light and the mixing is large, the most important input parameter is $m_A$, whereas ignorance of $\mu$ could still allow for a fair estimate of $m_{\tilde{t}}$
and $X_t$ in the stop sector. Moreover, we find that throughout the MSSM golden region the event rate $B\sigma$ is significantly smaller than the standard
model rate, which implies it may take more time to make the discovery. Given that there is no known method to directly measure 
the stop mixing parameter $X_t$, it will be important to combine the study presented here with the proposal in Ref.~\cite{Dermisek:2007fi} to get
indirect measurements on $m_{\tilde{t}}$ and $X_t$.

\section*{Acknowledgement}
This work was supported in part by U.S. DOE under contract 
DE-AC02-06CH11357. I. L. acknowledges valuable conversations with T. Tait and C. Wagner.

\end{document}